\documentclass[twocolumn,aps,pre,showpacs]{revtex4}
\usepackage[latin1]{inputenc}
\usepackage{graphicx}
\usepackage{latexsym}
\usepackage{amssymb}
\usepackage{amsmath}

\def\d{{\rm d}}

\def\u{{\bf u}}

\def\x{{\bf x}}
\def\T{{\boldsymbol{\sf T}}}
\def\S{{\boldsymbol{\sf S}}}
\def\R{{\boldsymbol{\sf R}}}

\def\F{{\bf F}}
\def\f{{\bf f}}
\def\U{{\bf U}}
\def\X{{\bf X}}

\def\bOmega{{\boldsymbol\Omega}}
\def\smalze{{\scriptscriptstyle{(0)}}}
\def\smalun{{\scriptscriptstyle{(1)}}}
\def\smaldu{{\scriptscriptstyle{(2)}}}

\def\beq{\begin{equation}}
\def\eeq{\end{equation}}
\begin{document}

\title{Passive swimming in low Reynolds number flows}
\author{Piero Olla}
\affiliation{ISAC-CNR and INFN, Sez. Cagliari, I--09042 Monserrato, Italy.}
\date{\today}

\begin{abstract}
The possibility of microscopic swimming by extraction of energy from an external flow
is discussed, focusing on the migration of a simple trimer across 
a linear shear flow.
The geometric properties of
swimming, together with the possible generalization to the case of a vesicle,
are analyzed.
The mechanism of energy extraction
from the flow appears to be the generalization to a discrete swimmer of the tank-treading 
regime of a vesicle. 
The swimmer takes advantage of the external flow by both
extracting energy for swimming and ''sailing'' through it.
The migration velocity is found to scale linearly in 
the stroke amplitude, and not quadratically as in a quiescent fluid.
This effect turns out to be connected with the non-applicability of the scallop
theorem in the presence of  external flow fields.


\end{abstract}

\maketitle
One of the problems that will have to be solved in
the realization of microscopic artificial swimmers (''microbots'')
is the energy source for locomotion. One possibility, 
is that of an external drive, e.g. by means of micromagnets inserted in the
device \cite{dreyfus05}. The final goal, however, is that of autonomous microbots
that are able to migrate, like their biologic companions, in  response
to external stimuli such as gradients in a chemical or temperature 
field.

Many situations involve swimming in a non-quiescent fluid, say the interior
of a blood vessel or the neighborhood of other swimming microorganisms.
These situations involve the presence of gradients in the fluid velocity
and the possibility, in principle, of energy conversion from external flow
to swimming.  A similar idea has been put
forward as regards the possibility of microscopic swimming aided by
rectification of thermal fluctuations \cite{golestanian09}.

Of course, unless imagining some sort of energy storage in the
microbot, the swimming velocity would be quite small, of the order of the
fluid velocity difference across the device body. Nevertheless, this
may be not an issue if long time or collective behaviors are 
sought.
Examples in nature of such behaviors do exist. A most notable one is
the Fahraeus-Lindwist effect: red blood cells in small arteries 
passively adapt their shape to the flow and
align in the middle of the vessel, thus decreasing
the vessel fluid-mechanic resistivity \cite{vand48}. 

Microscopic swimming takes places in an environment dominated by viscous
stresses, in which geometrical aspects play a dominant role. 
A strategy that has been
utilized to highlight these aspects, has been to focus 
on systems with a minimal number of degrees of freedom. 
Beyond various kinds of squirmers \cite{lighthill52}, 
examples are the three-sphere swimmer \cite{najafi04,pooley07},
the ''pushmepullyou'' \cite{avron05} and Purcell's three-link swimmer 
\cite{purcell77,becker03}.

It is an easy guess that some of the geometrical constrains
valid in a quiescent fluid, will cease to hold in the presence of
an external flow. In particular, the scallop theorem statement \cite{purcell77}, 
that reversing a sequence of deformations brings back the swimmer 
to its original position, 
will cease to be true.

\begin{figure}
\begin{center}
\includegraphics[draft=false,width=4.5cm]{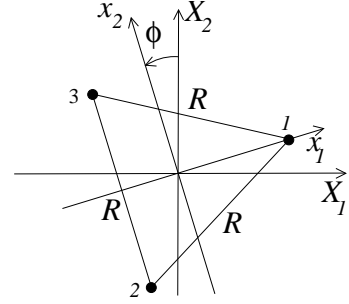}
\caption{
The coordinate system. The bead coordinates of the undeformed trimer 
in the rotating frame, are 
$\x^\smalze_1/R=(1/\sqrt{3},0,0)$,
$\x_2^\smalze/R=-(1/(2\sqrt{3}),1/2,0)$, $\x_3^\smalze/R=(-1/(2\sqrt{3}),1/2,0)$.
}
\label{orient}
\end{center}
\end{figure}

We are going to discuss an example of device that swims 
extracting energy out of an external flow, based on a 
generalization of the three-spheres swimmer of \cite{najafi04} (see Fig. \ref{orient}).
The swimmer is placed in an unbounded linear shear flow 
\beq
\bar\U(\X)= \bar\U(0)+(0,\alpha X_1,0),
\label{eq1}
\eeq
and wants to migrate along the gradient direction $X_1$.
%
%
We assume absence of external forces and torques on the trimer, 
except for those by the fluid, and neglect the interaction between the 
trimer arms and the fluid. 

To lowest order in the ratio $a/R$ between bead radii and arm lengths
(Stokeslet approximation \cite{happel}), 
the contribution to the trimer dynamics from the torque on the beads by 
the flow is disregarded. Faxen corrections at the bead scale and
higher order images from interaction of the flow perturbation with the
beads are disregarded as well. In this approximation,
the equation of motion for the trimer can be written
in the form
\beq
\dot\X_i=\bar\U(\X_i)+\tilde\U_i(t)+\F_i/\sigma,
\label{eq2}
\eeq
where $\X_i$ is the coordinate of the $i$-th bead, $\F_i$
is the force on bead $i$ by the rest of the trimer,
$\sigma=6\pi \mu a$, with $\mu$ the fluid viscosity, is the
Stokes drag, and $\tilde\U_i(t)$ is 
the flow perturbation generated in $\X_i$ by the other beads.

Contrary to the problem of the swimmer in a quiescent fluid, we see that the center of 
mass of the trimer $\bar\X=(\X_1+\X_2+\X_3)/3$
plays a special role, which allows, in the language of \cite{shapere89}, to fix
the gauge for the problem. In fact,
absence of external forces $\sum_i\F_i=0$, and linearity of the shear
imply, from Eq. (\ref{eq2}):
$\dot{\bar\X}=\bar\U(\bar\X)
+\sum_i\tilde\U_i(t)/3$. In the absence of flow perturbation, the trimer 
would move as a point tracer located at its center of mass.
Averaging over time the contribution by the flow perturbation gives the migration velocity
of the trimer:
\beq
\U^{migr}=(1/3)\sum_i\langle\tilde\U_i\rangle.
\label{eq3}
\eeq
Similar conclusions can be drawn as regards rotation around an axis passing through the swimmer 
center of mass. The rotation frequency $\bOmega$ can be obtained from Eq. (\ref{eq2}) imposing
absence of external torques: $\sum_i\X_i\times\F_i=0$, that gives 
$\bOmega=I^{-1}\sum_i\X_i\times[\bar\U(\X_i)+\tilde\U_i(t)]$,
$I=\sum_i|\X_i|^2$. Neglecting the effect of the flow perturbation $\tilde\U$, 
we find the constant rotation rate \cite{note}:
\beq
\bar\Omega^\smalze=\alpha/2,
\label{eq4}
\eeq 
that coincides with the flow vorticity
(the superscript indicates order in the deformation).
This a bonus from the point of view of calculation 
that would not be available for less symmetric shapes 
(for instance, in the Stokeslet approximation, 
a dimer would tend to align with the flow).

It is convenient to introduce a reference frame with origin in the center of mass,
rotating solidly with the trimer, as illustrated in Fig. \ref{orient}. 
Equation (\ref{eq4}) allows to convert time averages into angular averages:
\beq
\langle h\rangle=\frac{1}{2\pi}
\int_0^{2\pi}\d\phi\Big[1+\frac{2}{\alpha}\bar\Omega^\smalun(\phi)+\ldots\Big]h(\phi),
\label{merda}
\eeq
with $\phi$ the angle between the two reference frames (see Fig. \ref{orient}).
We shall use small case to identify vectors in the rotating frame.
Equation (\ref{eq2}) will
become in this frame an evolution equation for the trimer internal degrees of freedom. 

The flow perturbation produced by the trimer can be expressed in function of the
forces on the beads, and, thanks to linearity of low Reynolds number hydrodynamics, the
resulting relation will be linear as well; in the rotating reference frame:
\beq
\tilde\u_i(t)=\sum_{j\ne i}\T(\x_i,\x_j)\f_j,
\label{eq5}
\eeq
where 
$\T$ is the off-diagonal part of the Oseen tensor, relating, in the absence 
of an external flow,
the $\dot\x_i$'s and the $\f_j$'s \cite{happel}. 
To lowest order in $a/R$,
we can then write $\T(\x_i,\x_j)\simeq \T(\x_{ij})$, $\x_{ij}\equiv \x_i-\x_j$,
where, for $i\ne j$ \cite{happel}:
\beq
\T(\x_{ij})
=\frac{3a}{4\sigma}\Big[\frac{{\bf 1}}{|\x_{ij}|}+\frac{\x_{ij}\x_{ij}}
{|\x_{ij}|^3}\Big].
\label{eq6}
\eeq
Substituting Eq. (\ref{eq5}) into Eq. (\ref{eq3}) and assuming exchange symmetry 
between the beads in the trimer:
\beq
\U^{migr}=\langle\R\T_1\f_1\rangle,
\label{eq7}
\eeq
where $\T_1=\T(\x_{21})+\T(\x_{31})$ and 
$\R$, with $R_{11}=R_{22}=\cos\phi$, $R_{21}=-R_{12}=\sin\phi$, 
is the rotation matrix back to the fixed frame.

It is easy to see that a rigid trimer, due to the symmetry of its shape 
and of the external flow, would be unable to migrate.
We must consider deformations.

\begin{figure}
\begin{center}
\includegraphics[draft=false,height=3.5cm]{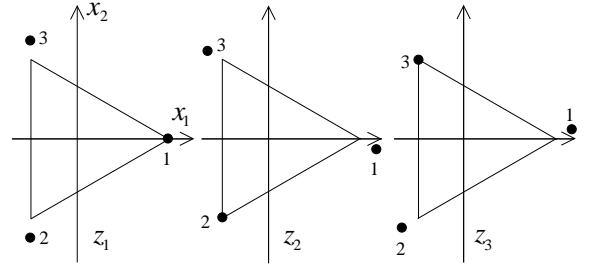}
\caption{
Deformations of the trimer corresponding to $z_i>0$ for $i=1,2,3$. In the three cases
$z_j=0$ for $j\ne i$.
}
\label{deform}
\end{center}
\end{figure}
%
%
We can express the trimer deformation in terms of three independent parameters
$z_1,z_2,z_3$ as follows:
\begin{eqnarray}
\x^\smalun_1/R&=&((\sqrt{3}/2)(z_2+z_3),(z_3-z_2)/2,0),
\nonumber
\\
\x^\smalun_2/R&=&(-(\sqrt{3}/2)z_3,-z_1-z_3/2,0),
\label{eq8}
\\
\x^\smalun_3/R&=&(-(\sqrt{3}/2)z_2,z_1+z_2/2,0).
\nonumber
\end{eqnarray}
As illustrated in Fig. \ref{deform}, positive $z_i$ corresponds to stretching
of the arm opposite to bead $i$.
From Eq. (\ref{eq7}), to determine $\U^{migr}$, we must calculate $\f_1$ at least to
$O(z)$, $z=\langle\sum_iz_i^2\rangle^{1/2}$, 
but  we can neglect the effect of the field perturbation.
The rotating frame version of Eq. (\ref{eq2}) reads at $O((a/R)^0)$:
$\f_i/\sigma=\dot\x_i-\bar\u(\x_i)$. 

At $O(z^0)$ there are no deformations and $\dot\x^\smalze=0$.
To next order, from Eq. (\ref{merda}), we can write $\dot\x^\smalun=(\alpha/2){\x^\smalun}'$ with
prime indicating $\d/\d\phi$. From Eq. (\ref{eq4}),
the rotation velocity equals to $O(z^0)$ the vorticity of the field; hence
$\bar\u^\smalze$ is purely due to strain, $\bar\u^\smalze(\x)=\bar\S\x$, with
\beq
\bar S_{11}=-\bar S_{22}=\frac{\alpha}{2}\sin 2\phi,
\ 
\bar S_{12}=\bar S_{21}=\frac{\alpha}{2}\cos 2\phi.
\label{eq9}
\eeq
At $O(z)$, $\bar\u$ receives a vorticity components from the correction to the trimer 
rotation velocity: $\bar\u^\smalun(\x)=-\bar\bOmega^\smalun\times\x$. 
To $O(z)$ the trimer evolution equation will read therefore
\begin{eqnarray}
\f^\smalze_i&=&-\sigma\bar\S\x_i^\smalze,
\label{eq10}
\\
\f^\smalun_i&=&\sigma[(\alpha/2){\x^\smalun_i}'-\bar\S\x_i^\smalun
+\bar\bOmega^\smalun\times\x_i^\smalze].
\label{eq11}
\end{eqnarray}
To obtain $\bar\Omega^\smalun$, we apply $\x^\smalze_i\times$ at both sides of Eq. (\ref{eq11}) 
and sum over $i$, with the result
$\bar\bOmega^\smalun=\sum_i\x_i^\smalze\times\bar\S\x_i^\smalun/I^\smalze$,
$I^\smalze=\sum_i|\x^\smalze_i|^2=R^2$. Direct calculation from Eqs. (\ref{eq8}-\ref{eq9})
gives then
\beq
\bar\Omega^\smalun=\Big[\sqrt{3}\,(z_2-z_3)\sin 2\phi
+(z_2+z_3-2z_1)\cos 2\phi\Big]\frac{\alpha}{4}.
\label{eq12}
\eeq

\begin{figure}
\begin{center}
\includegraphics[draft=false,width=6.2cm]{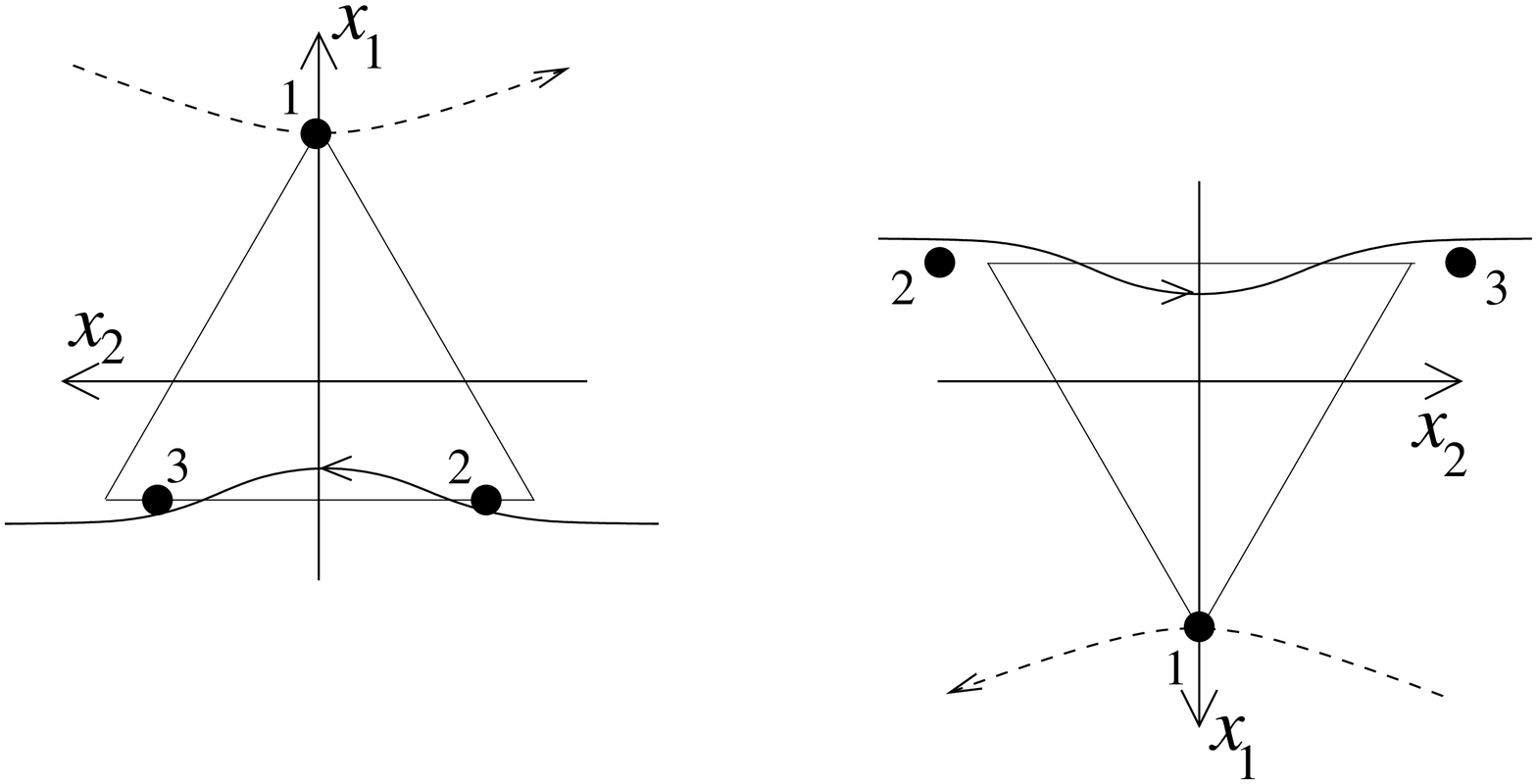}
\caption{The swimming strategy. 
When $\phi=\pi/2$ (left), arm $23$ contracts, which turns out to maximize the
drag along $X_1$ by bead 1's Stokeslet field [continuous line; see Eq. (\ref{eq6})].
The opposite occurs for $\phi=-\pi/2$ (right).
Dashed lines identify the strain component of $\bar\U$.
}
\label{nuoto}
\end{center}
\end{figure}

Substituting Eqs. (\ref{merda},\ref{eq6})
and (\ref{eq8}-\ref{eq12}) into Eq. (\ref{eq7}), we find the lowest order contribution 
$\U^{migr}\simeq\langle \R(\T_1^\smalze\f_1^\smalun
+\T_1^\smalun\f_1^\smalze)\rangle^\smalze$, where $\langle .\rangle^\smalze$ 
indicates the uniform part of the average in Eq. 
(\ref{merda}). 
Imposing exchange symmetry among the beads:
\beq
z_i=\sum_n[A_n\cos n\phi_i+B_n\sin n\phi_i],
\label{eq13}
\eeq
where $\phi_1=\phi$, $\phi_{2,3}=\phi\mp 2\pi/3$. 
We see that Eqs. (\ref{merda},\ref{eq9},\ref{eq12}) bring into
Eq. (\ref{eq7}) factors $\cos 2\phi$ and $\sin 2\phi$ that combined with the $\cos\phi$ and
$\sin\phi$ in $\R$, require, for a non zero migration velocity, 
contributions $\propto\cos\phi,\cos 3\phi,\sin\phi,\sin 3\phi$ in 
the deformation. 
Direct calculation, using
$T_1^{11}\simeq \beta\{7/2-[13z_1+29(z_2+z_3)]/8\}$, $T_1^{12}=T_1^{21}\simeq 
\beta(\sqrt{3}/8)(z_2-z_3)$ and
$\tilde T_1^{22}\simeq \beta\{5/2+[2z_1-31(z_2+z_3)]/\}8$, 
$\beta=3a/(4\sigma R)$, together with Eq. (\ref{eq13}),
gives in fact the result:
\beq
U_1^{migr}\simeq -\frac{\sqrt{3}\alpha a}{256}\Big[73B_1+13B_3\Big].
\label{eq14}
\eeq
The component $B_1$ corresponds to the pulsation regime depicted in Fig. \ref{nuoto}.
From here, the swimming strategy could be optimized minimizing at fixed $U^{migr}_1$ 
the deformation amplitude $z$ or some more appropriate cost function \cite{avron04}.

Notice that, of all the contributions to migration, perhaps, only the 
one coming from the first term to right hand side of Eq. (\ref{eq11}) qualifies
as real swimming. It is in fact the only direct contribution from deformation
to bead movement with respect to the fluid. All the others could be loosely 
described as a sort of sailing, in which the swimmer shape adapts to catch
the external flow.

The way the swimmer actually swims, is illustrated in Fig. \ref{nuoto}.
We see that propulsion does not come from direct drag by $\bar\U$, rather,
from drag 
by the perturbation $\tilde\U$ generated by interaction between the trimer
and the external flow.
Imposing exchange symmetry among the beads, allows, 
on the same line of Eq. (\ref{eq7}), to focus on the Stokeslet field
of a single bead, say bead 1. 
%
%
%
%
Let us consider the contribution $B_1\sin\phi_1$.
We see that the Stokeslet field generated by bead 1 in response to the strain component 
of $\bar\U$,
has a positive or negative component at beads $2,3$, depending on whether
$0<\phi<\pi$ or $\pi<\phi<2\pi$. Migration is produced by the deformation induced 
symmetry breaking between the two orientations.

The swimmer of Eq. (\ref{eq13}) behaves as a small engine, with the three arms stretching
and contracting in sequence. From linearity in $z$ of $U^{migr}_1$, however, 
migration in the symmetric case of Eq. (\ref{eq13}), 
implies migration also in the presence of 
a single deformable link.
The scallop theorem
does not apply in the present situation: 
the swimmer orientation and position are determined jointly by the past history of
deformations and of the gradients of $\bar\U$ experienced: even correcting
for the trivial displacement by $\bar\U$, 
reversing the deformations only, would not be sufficient to bring 
the trimer back to its original position. Similar ''violations'' of the scallop theorem
have been observed in \cite{alexander08}, where the role of external flow was played
by the perturbation generated in the fluid by other swimmers.

This situation has a counterpart in the fact that swimming in an external flow
occurs already at $O(z)$, and not at $O(z^2)$. The swimmer achieves this,
exploiting 
the external flow to change orientation between reversed strokes;
in a quiescent fluid, the orientation change would require 
other stroke components, and this is responsible for $O(z^2)$ 
swimming efficiency \cite{shapere89}.

It is actually possible for the swimmer to extract the energy needed for migration 
directly from the external flow.
The average power
exerted by the swimmer on the fluid is $\dot W=3\langle\dot\x_1\cdot\f_1\rangle$.
Now, from $\dot\x^\smalze_i=0$, the first contribution is $\dot W^\smalun=
3\langle\dot\x^\smalun_1\cdot\f^\smalze_1\rangle^\smalze$, but the $\x^\smalun$ components 
contributing
to swimming, entering Eq. (\ref{eq14}), are odd in $\phi$, while $\f^\smalze_i$, from Eq. (\ref{eq10}), 
is even. The same occurs, from Eqs. (\ref{merda}) and (\ref{eq12}), with $\langle.\rangle^\smalun$.
Assuming that $\x^\smalun$ receives contributions only from $B_{1,3}$,
the lowest order contribution to the power will be
\beq
\dot W^\smaldu=3[\langle\dot\x_1^\smalun\cdot\f_1^\smalun\rangle^\smalze+
\langle\dot\x_1^\smaldu\cdot\f_1^\smalze\rangle^\smalze];
\label{eq15}
\eeq
we see that choosing $\x^\smaldu$ in appropriate way, we could set
$W^\smaldu=0$, compensating to $O(z^2)$ the energy lost in swimming.
From Eq. (\ref{eq10}), the only terms in $\x^\smaldu$ contributing to 
$\dot W^\smaldu$ are
$\propto\cos 2\phi,\sin 2\phi$. 
Substituting Eqs. (\ref{eq8},\ref{eq10},\ref{eq13}) into
Eq. (\ref{eq15}) we obtain:
\beq
A_2>\frac{3\sqrt{3}}{32}(B_1^2-5B_1B_3+36B_3^2),
\label{eq16}
\eeq
where the inequality accounts for the internal dissipation of the device
[at this point the condition that $A_2=O(z^2)$ becomes irrelevant].
Looking at Fig. \ref{nuoto} and focusing again on bead 1, we see that this deformation 
corresponds to 
arm 23 being stretched when it is parallel to the flow direction, and contracted when
it is perpendicular.
Energy extraction from the flow comes from the fact that the $|\x_{23}|$ grows in the stretching
quadrants of the external strain, and decreases in the contracting ones 
(see Fig. \ref{nuoto}).

\begin{figure}
\begin{center}
\includegraphics[draft=false,width=6.2cm]{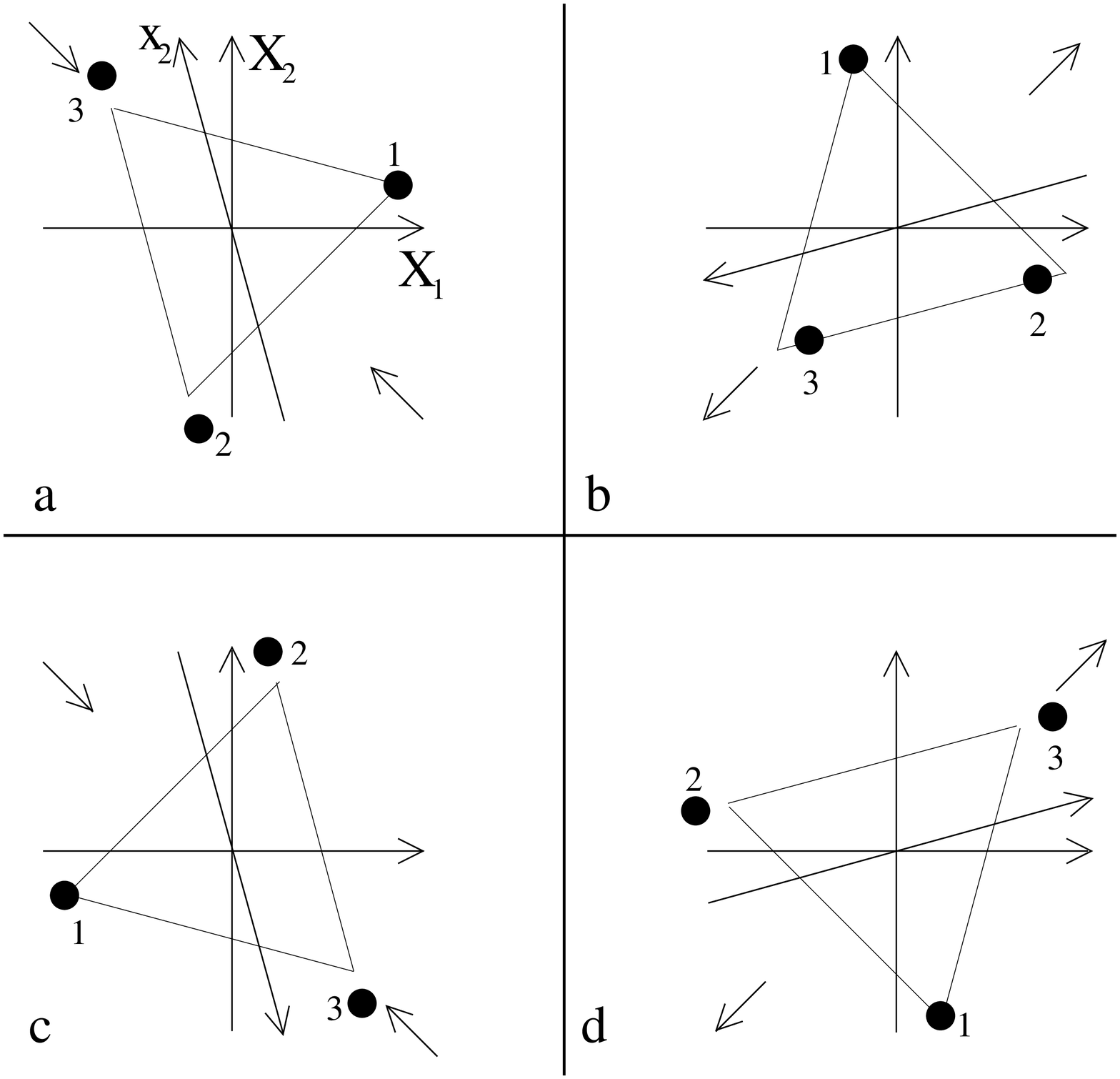}
\caption{Swimming from braking. Suppose $\x_{23}$ enters the contracting
quadrants in a stretched state $(a)$. The brake is relased and the arm contracts, till it
enters the stretching quadrants fully contracted $(b)$. The brake remains off till the arm 
returns, now fully stretched, into the contracting quadrants $(c)$. At this point, the brake 
is acted on and $|\x_{23}|$ remains stretched until, passing through $(d)$, 
it returns to $(a)$.
}
\label{brake}
\end{center}
\end{figure}

What mechanism could allow a three-sphere swimmer, such as the one described, to
passively undergo the deformation sequence depicted in Eqs. (\ref{eq14}) 
and (\ref{eq16})? A possibility is illustrated in Fig. \ref{brake}: a system
of orientation dependent brakes, controlling the stretching and contraction 
of the trimer arms in response to the external strain (a more quantitative 
description of this mechanism is provided in \cite{olla10}).
This mechanism contains all the
required aspects of conversion of external flow to deformation and then to swim.
We notice the asymmetry between the two configurations 
$(b)$ and $(d)$, that is necessary to migration. At the same time, 
comparing orientations $(a)$ and $(c)$ with $(b)$ and $(d)$, 
we see that arm 23 is in the stretched state when aligned with $\bar\U$, 
meaning that
energy is extracted from the external flow.
\vskip 5pt
All the behaviors that have been described, derive from
geometrical properties of the swimmer and could be expected to remain valid
away from the small deformation regime considered.
More importantly, it should be possible to construct a continuous counterpart
of the trimer, for which the
migration velocity predictions of Eq. (\ref{eq14}) remain qualitatively valid,
albeit substitution between the typical size of the moving parts in the two 
situations: $a\to R$. Such a continuous swimmer should have the properties
imposed by Eqs. (\ref{eq14}) and (\ref{eq16}): an overall shape in the 
laboratory frame that is stretched along $\bar\U$ and asymmetric; say an egg 
with the tip pointing at positive $X_2$. In other words, a tank-treading
cell \cite{keller82,coupier08} with a fixed orientation asymmetric shape.
Unfortunately, this would require 
an orientation sensitive control of the device structure 
(compare with Fig. \ref{brake}), and the practical realization 
of such a device at the microscale would be very difficult.

Nevertheless, from a conceptual point of view, some complex version of
tank-treading appears to be the most effective strategy for a swimmer
to exploit the presence of an external flow. A linear swimmer, say, could very well
extract energy from $\bar\U$ in a tumbling cycle, 
and use it to generate swimming strokes as if in a quiescent fluid.
However, beyond having a lower stroke frequency than a comparable triangular device
(it would spend most of the time aligned with the flow), it would have only quadratic
efficiency in the stroke amplitude.


\begin{thebibliography}{10}
\bibitem{note} 
It can be shown that a device formed by four beads at the vertices of a regular 
tetrahedron will share this property, irrespective of its initial orientation.
\end{thebibliography}
\end{document}